\documentclass[twocolumn, amsmath, amssymb, aps, prapplied]{revtex4-2}
\usepackage[utf8]{inputenc}
\usepackage[T1]{fontenc}

\usepackage{dcolumn}

\usepackage{csquotes}

\usepackage{mathtools}
\usepackage{bm}
\usepackage{physics}

\usepackage{booktabs}

\usepackage{graphicx} 

\usepackage{url}

\usepackage{hyperref}

\usepackage{cleveref}
\crefname{equation}{Eq.}{Eqs.}
\crefname{section}{Sec.}{Secs.}
\crefname{table}{Table}{Tables}
\crefname{figure}{Fig.}{Figs.}

\usepackage{xcolor}

\NewDocumentCommand{\ac}{}{a_\mathrm{c}}

\NewDocumentEnvironment{chblock}{}{\color{red}}{}
\usepackage[normalem]{ulem} 

\DeclareMathOperator*{\argmin}{arg\,min}
\DeclareMathOperator{\sign}{sign}
\DeclareMathOperator{\lmax}{\lambda_\mathrm{max}}

\begin{document}

\title{Soft vector spins with dimensional annealing for combinatorial optimization}

\author{Marvin Syed}
\email{marvin.syed@strath.ac.uk}
\affiliation{%
 Department of Applied Mathematics and Theoretical Physics, University of Cambridge\\
 Wilberforce Road, Cambridge CB3~0WA, United Kingdom%
}
\affiliation{%
 Department of Physics and SUPA, University of Strathclyde, Glasgow G4 0NG, United Kingdom%
}

\author{Richard Zhipeng Wang}
\affiliation{%
 Department of Applied Mathematics and Theoretical Physics, University of Cambridge\\
 Wilberforce Road, Cambridge CB3~0WA, United Kingdom%
}

\author{Natalia G. Berloff}
\affiliation{%
 Department of Applied Mathematics and Theoretical Physics, University of Cambridge\\
 Wilberforce Road, Cambridge CB3~0WA, United Kingdom%
}

\date{\today}

\begin{abstract}
	Recently, purpose-built analog hardware that can efficiently minimize the
	Ising energy and thereby solve a variety of combinatorial optimization
	problems has been receiving widespread attention.
	In this work, we show how multidimensional, vectorial degrees of freedom,
	that are either naturally present or can be artificially created in such
	hardware, could strengthen the capability to find optimal solutions to
	optimization problems.
	In order to achieve this, we introduce a simple model of soft vector spins
	that should be implementable on a variety of analog hardware platforms as
	well as three different dimensional annealing methods which harness the
	enlarged phase space of the vectorial degrees of freedom to minimize the
	Ising energy.
	We perform simulations on different benchmark problems and show that for
	all dimensional annealing methods we tested, vectorial degrees of freedom
	improve upon one-dimensional degrees of freedom when it comes to finding
	the ground state of the Ising model.
	In particular, we find that this advantage becomes most pronounced for $d \gtrsim 3$
	dimensional degrees of freedom, with diminishing returns as the dimension is increased
	further.
	Our results could inspire new analog optimization hardware and algorithms
	that explicitly incorporate the advantage of vectorial degrees of freedom.
\end{abstract}

\maketitle

\section{Introduction}
Solving combinatorial optimization problems is of interest in a variety of
fields.
Many of these problems can be efficiently represented in terms of a classical
Ising model~\cite{lucasIsingFormulationsMany2014}, such that the variables
become Ising spins, and the objective becomes finding the minimum of the Ising
energy, that is, finding
\begin{equation}
    \argmin_{\bm{s} \in \{-1, +1\}^N} -\frac{1}{2} \sum_{i,j=1}^N J_{ij} s_i s_j \, ,
\end{equation}
where $J$ is a symmetric matrix.
In recent years, this fact has sparked considerable interest in developing
purpose-built hardware aimed at minimizing the Ising energy.
While some works in this area make use of conventional digital computing
hardware (
	for example, field programmable gate arrays (FPGAs)~\cite{%
		goto2019CombinatorialOptimizationSimulating,
		kanao2022SimulatedBifurcationHigherorder%
	}, or other custom CMOS
	hardware~\cite{aramon2019PhysicsInspiredOptimizationQuadratic}
),
many proposed hardware platforms are based on analog physical systems.
These include ultracold atoms~\cite{byrnes2011AcceleratedOptimizationProblem},
injection-locked lasers~\cite{utsunomiya2011MappingIsingModels},
degenerate optical parametric oscillators~\cite{%
	wang2013CoherentIsingMachine,
	yamamoto2017CoherentIsingMachines%
},
polariton condensates~\cite{%
	berloff2017RealizingClassicalXY,
	lagoudakis2017PolaritonGraphSimulator,
	kalinin2018SimulatingIsing,
	kalinin2018global%
},
spatially modulated light~\cite{
	veraldi2025FullyProgrammableSpatial,
	wang2025EfficientComputationUsing%
},
polarization oscillators~\cite{chiavazzo2025IsingMachineDimensional},
or electronic oscillators~\cite{%
	wang2019OIMOscillatorBasedIsing,
	albertsson2023HighlyReconfigurableOscillatorbased%
}.
Often, the analog systems can be described as a dynamical system of interacting
analog units of the form
\begin{equation}\label{eq:softising}
	\dv{t} x_i = f_i(x_1, \dots, x_N, t) \, .
\end{equation}
where \(f_i\) describes some (possibly time-dependent)
forces~\cite{syedPhysicsEnhancedBifurcationOptimisers2023}.
We will refer to the analog units \(x_i \in \mathbb{R}\) as \emph{soft (Ising)
spins}.
During operation, the soft spins evolve according to \cref{eq:softising} for a
specified amount of time or until some convergence criterion is reached.
Afterwards, the soft spins are read out and transformed into \enquote{hard}
Ising spins via \(s_i = \sign(x_i)\).

\begin{figure}
	\centering
	\includegraphics[width=\linewidth]{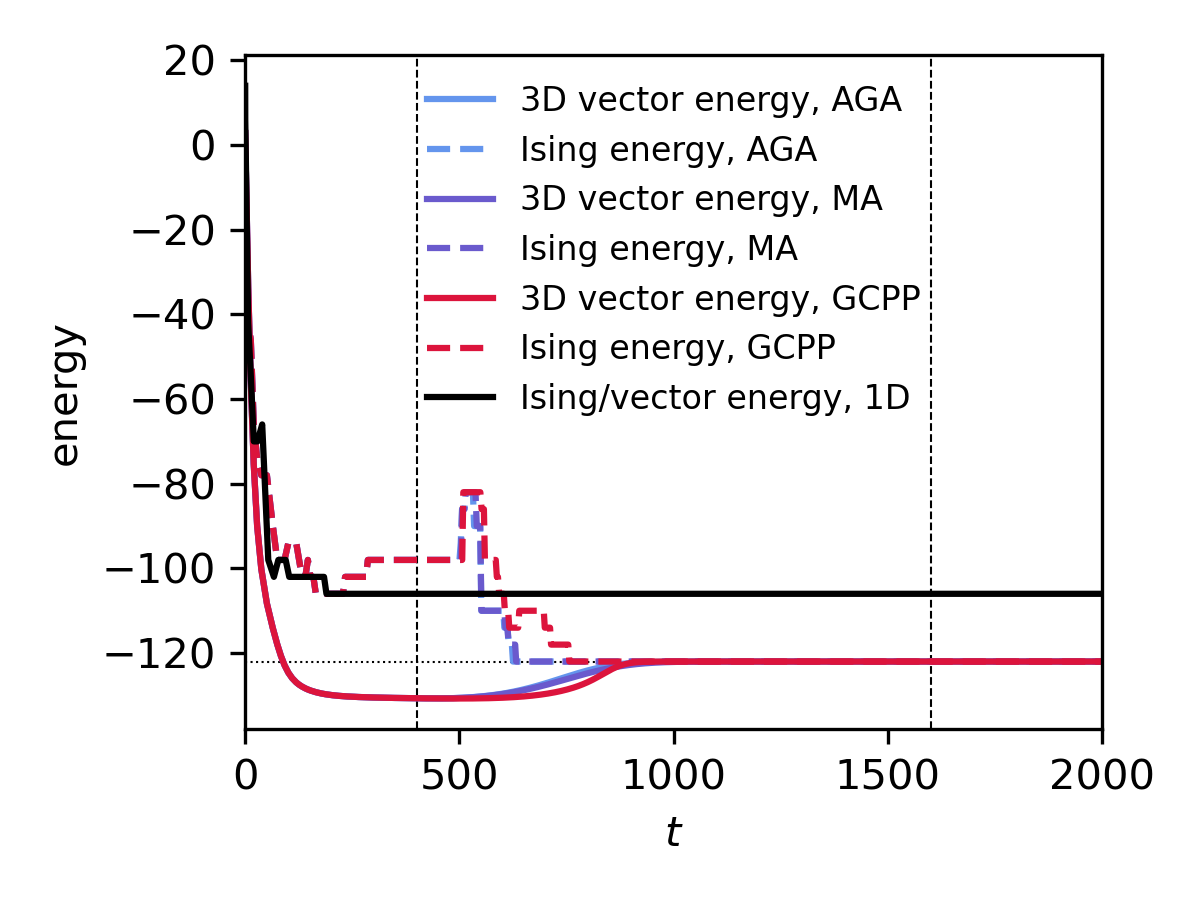}
	\caption{%
		Example of \(d=3\) soft vector spins traversing the Ising energy landscape.
        For each point in time we evaluate the vector spin energy, $E_\mathrm{vec}$, via $\bm{s}_i = \bm{x}_i / \norm{\bm{x}_i}$, and the Ising energy, $E_\mathrm{Ising}$, via the method in \cref{sub:proj}.
		Solid colored lines show the resulting trajectories for
		\(E_{\mathrm{vec}}\) using different dimensional annealing
		methods (described in detail later in the text); the dashed lines show the ones for \(E_{\mathrm{Ising}}\).
		For comparison, the evolution of a \(d=1\) soft Ising spin system is
		shown in black (in one dimension \(E_{\mathrm{vec}}\) and
		\(E_{\mathrm{Ising}}\) coincide).
		Vertical dashed lines show the time period where dimensional annealing
		is active, while the dotted horizontal line shows the Ising ground
		state energy.
		Within this region, the dimensional annealing process enables jumps
		over Ising energy barriers that are not possible in one dimension.%
	}
	\label{fig:en_evol}
\end{figure}

In this work, we introduce a multi-dimensional generalization of soft Ising
spins,
\emph{soft vector spins}, that evolve according to an equation of the form
\begin{equation}\label{eq:softvec}
	\dv{t} \bm{x}_i = \bm{f}_i(\bm{x}_1, \dots, \bm{x}_N, t) \, ,
\end{equation}
where the soft vector spins, \(\bm{x}_i \in \mathbb{R}^d\), are now vectorial
degrees of freedom.
This is motivated by a couple of different reasons.
While the ultimate aim is to minimize the Ising energy, many hardware platforms
naturally operate in a higher-dimensional state space with variables possessing
a continuous phase degree of freedom, thus making them naturally more suited to
minimize the energy of vector spin models like the XY model.
Although minimizing vector spin models has applications in itself~\cite{%
	kim2024CombinatorialClusteringCoherent,
	wang2025PhaseRetrievalGainBased%
}, the solution of the Ising problem is often still desired.
Because of this, many approaches include a mechanism that biases dynamics in
favor of aligning all spins along a single axis, for example through
second-harmonic injection locking in polariton
networks~\cite{kalinin2018global} and oscillator Ising
machines~\cite{albertsson2023HighlyReconfigurableOscillatorbased}.
Beyond being a hardware necessity, recent works have employed different such
\emph{dimensionality reducing} mechanisms to obtain better performing
optimization methods~\cite{%
	calvanesestrinati2022MultidimensionalHyperspinMachine,
	calvanesestrinati2024HyperscalingCoherentHyperspin,
	cummins2025VectorIsingSpin%
}.
This is done by varying the strength of the dimensionality reducing mechanism
during time-evolution.
In accordance with Refs.~\cite{%
	calvanesestrinati2022MultidimensionalHyperspinMachine,
	calvanesestrinati2024HyperscalingCoherentHyperspin%
}, we will refer to this process as \emph{dimensional annealing}.
Intuitively, the extra degrees of freedom of soft vector spins turn local
minima into saddle points opening up new \enquote{escape routes} during the
dimensional annealing process, thereby increasing the likelihood of finding the
global minimum (see \cref{fig:en_evol}).
On the other hand, there is a trade-off in the fact that dimensional reduction
mechanisms can distort the energy landscape and make it harder to identify the
true global minimum.

Our aim in this work is to compare dimensional annealing with a selection of
dimensional reduction methods on a set of Ising benchmark problems, and study
their dependence on the dimensionality of the soft vector spin variables.
In order to do so, we start with a base model of coupled soft vector spins
which can be used to describe a variety of concrete hardware platforms.
There are a variety of ways to implement dimensionality reducing mechanisms in
such a model.
We focus on three:
\begin{enumerate}
	\item An anisotropic gain, where there is a stronger gain for one specific
		component of the vector spins.
		We show that this is essentially the same as second-harmonic
		injection-locking which was considered in
		Refs.~\cite{kalinin2018global,albertsson2023HighlyReconfigurableOscillatorbased},
		for example.
	\item An anisotropic metric, where the couplings among spins are stronger
		for a specific vector component.
		This mechanism has previously been discussed in the context of the
		\emph{Hyperspin
		Machine}~\cite{calvanesestrinati2022MultidimensionalHyperspinMachine,calvanesestrinati2024HyperscalingCoherentHyperspin}.
	\item A generalized cross-product penalty.
		Such a penalty term was introduced for the \emph{Vector Ising Spin
		Annealer} (VISA)~\cite{cummins2025VectorIsingSpin}, but only for three
		dimensional vector spins.
		We show that this type of penalty can be implemented for any dimension.
\end{enumerate}
Our findings show that soft vector spins with dimensional annealing have an
increased or equal probability of finding the ground state compared to soft
Ising spins in all test cases we considered.
Furthermore, higher dimensions \(d \geq 2\) generally fare better for all three dimensional
reduction methods, with diminishing returns for $d \geq 3$.
These results imply that existing and future analog optimization hardware could
benefit from making use of analog units with vectorial degrees of freedom.

The outline of the paper is as follows:
In Sec.~\ref{sec:model}, we introduce and describe the elementary properties of
the base model of coupled soft vector spins and describe how to add dimensional
reduction mechanisms to it.
Sec.~\ref{sec:sims} contains the methodology and results of our numerical
simulations.
We conclude with a discussion of our results in Sec.~\ref{sec:disc}.

\section{Model}\label{sec:model}
\subsection{Base model}
We begin with a base model of a dynamical system consisting of \(N\) coupled
soft vector spins evolving via
\begin{equation} \label{eq:softspindyn}
    \dv{t} \bm{x}_i = a_i(t) \bm{x}_i - \norm{\bm{x}_i}^2 \bm{x}_i + \sum_{j=1}^N J_{ij} \bm{x}_j \, ,
\end{equation}
where \(\bm{x}_i \in \mathbb{R}^d\), $J$ is a symmetric coupling matrix with \(J_{ii}=0\),
and $a_i(t)$ is the time-dependent linear gain which is usually
increased (annealed) with time from an initial small value.
For \(d=1\) and \(a_i(t) = a(t)\), this reduces to the dynamics of the
\emph{Coherent Ising Machine} (CIM)~\cite{
    wang2013CoherentIsingMachine,%
    yamamoto2017CoherentIsingMachines%
}:
\begin{equation}
    \dv{t} x_i = a(t) x_i - x_i^3 + \sum_{j=1}^N J_{ij} x_j \, ,
\end{equation}
while for \(d=2\) we can use complex variables \(\psi_i\) to express the
equation of motion as
\begin{equation}
    \dv{t} \psi_i = a(t) \psi_i - \abs{\psi_i}^2 \psi_i + \sum_{j=1}^N J_{ij} \psi_j \, .
\end{equation}
Such equations frequently arise in systems used for gain-based
computing~\cite{cummins2025VectorIsingSpin} such as polariton
condensates~\cite{kalinin2018global}, and are an example of a wider class of
Andronov-Hopf oscillators systems that are common in analog optimization
hardware~\cite{syedPhysicsEnhancedBifurcationOptimisers2023}.

The soft vector spin system \eqref{eq:softspindyn} has an associated energy
landscape
at each point in time, s.t.,
\begin{equation}
	\dv{t} \vb{x} = -\vb{\nabla} E(\vb{x}, \vb{a}(t)) \, .
\end{equation}
For \(\vb{a}(t) = \vb{a} = (a_1, \dots, a_N)\), the energy function is
\begin{equation}\label{eq:energy}
	E(\vb{x}, \vb{a}) = \sum_i V(\bm{x}_i, a_i) - \frac{1}{2} \sum_{ij} J_{ij} \bm{x}_i \vdot \bm{x}_j \, ,
\end{equation}
where $\vb{x} = \bigl( \bm{x}_1, \dots, \bm{x}_N \bigr)$ and
\begin{equation}
    V(\bm{x}, a) =  -\frac{a}{2} \norm{\bm{x}}^2  + \frac{1}{4} \norm{\bm{x}}^4  \, .
\end{equation}
This energy function can be seen as a generalization of the energy landscape of
the CIM studied in Refs.~\cite{%
	yamamura2024GeometricLandscapeAnnealing,
	ghimenti2025GeometryDynamicsAnnealed%
} which represents the \(d=1\) case of the soft vector spin model.
In the past, similar models (also with \(d=1\)) were used in the context of
structural glasses~\cite{%
	kuhn1997RandomMatrixApproach,
	rainone2021MeanfieldModelInteracting%
}.

If the spins in \cref{eq:energy} were \enquote{hard} vector spins with
$\bm{x}_i \in \mathbb{S}^{d-1}$ living on the unit sphere, the second term in
Eq.~\eqref{eq:energy} would simply be the energy of a classical system of
vector spins.
The first term takes the form of the classic \enquote{sombrero} potential
acting on each spin individually.
This kind of potential often arises in systems exhibiting spontaneous symmetry
breaking.
In our setting we may look at it simply as a penalty that punishes soft spin configurations
with vector spins that stray too far from the origin (when $a < 0$) or the surface of the
sphere $\mathbb{S}^{d-1}(\sqrt{a})$ of radius $\sqrt{a}$ (when $a \geq 0$).
In particular, the potential well becomes deeper as the gain $a$ is increased,
ultimately leading to the soft vector spins converging towards hard vector
spins constrained to the sphere $\mathbb{S}^{d-1}(\sqrt{a})$.
This intuition can be put into more quantitative terms by examining the
low-gain and high-gain limit.

\subsubsection{Low-gain limit}
\label{sub:lowgain}
In this section, we assume homogeneous gains, \(a_i(t) = a(t)\) for all \(i\).
A similar picture, however, holds also for heterogeneous gains.

At low enough gain, the amplitudes of the soft spins will be driven close to
zero, i.e, $\norm{\bm{x}_i} \ll 1$.
This allows us to linearize \cref{eq:softspindyn}, yielding
\begin{equation}\label{eq:linearized}
    \dv{t} \bm{x}_i = \sum_j (a\delta_{ij} + J_{ij}) \bm{x}_j \, .
\end{equation}
The linearized system's behavior depends only on the eigenvalues of the matrix
$aI + J$ with the modes corresponding to positive (negative) eigenvalues
exponentially growing (decaying).
A transition occurs at the point $\ac = -\lmax(J)$, where the gain equals the
negative maximum eigenvalue of $J$.
Below $\ac$, all modes decay and the origin is the only fixed point of
\cref{eq:softspindyn}.
Above $\ac$, the fixed point at the origin becomes unstable and the amplitudes
start to grow causing the system to move away from the linear regime.
Close to the transition point, the soft spins align themselves in the direction
of the eigenvector associated to $\lmax(J)$.
More precisely, treating $a$ as quasi-static near the transition point, the linearized
equation admits the formal solution
\begin{equation}\label{eq:xlin}
    \vb{x}(t)
    = \exp(at) \exp((J \otimes I_d)t) \vb{x}(0) \, ,
\end{equation}
where $I_d$ is the $d \times d$ identity matrix, and $\otimes$ denotes the tensor product.
The exponentially fastest growing mode in \cref{eq:xlin} is given by the maximal eigenvector of $J \otimes I_d$.
Written in components, this is $x_i^\mu \propto
v_i e^\mu$, where $v_i$ are the components of the eigenvector associated to $\lmax(J)$ and $\bm{e}$ is
an arbitrary vector giving the direction of the spin vectors.
This means that in the vicinity of the transition point, the system finds a minimizer of the energy
\begin{equation}
    E_\mathrm{spec} = -\frac{1}{2} \sum_{ij} J_{ij} u_i u_j \, ,
\end{equation}
over all $u_i$ such that $\sum_i u_i^2 = 1$, which is a spectral approximation
of the problem of minimizing the Ising
energy~\cite{kalininComputationalComplexityContinuum2022,
yamamura2024GeometricLandscapeAnnealing},
\begin{equation}
    E_\mathrm{Ising} = -\frac{1}{2} \sum_{ij} J_{ij} s_i s_j \, .
\end{equation}
As the gain is increased further beyond the transition point, other
eigenvectors start to contribute more and the system goes beyond simply finding
the spectral approximation.

\subsubsection{High-gain limit}
Let us now examine the high-gain limit where $a \gg 1$.
In that case, $V(\bm{x}, a)$ has deep minima on the sphere where $\norm{\bm{x}}
= \sqrt{a}$.
This inspires us to make the change of variables $\bm{y} = a^{-1/2} \bm{x}$.
As a result, the energy expressed in the new variables has the form
\begin{equation}
    a^{-2}E = \sum_i \left(
        -\frac{1}{2} \norm{\bm{y}_i}^2
        + \frac{1}{4} \norm{\bm{y}_i}^4
    \right)
    - \frac{1}{2a} \sum_{ij} J_{ij} \bm{y}_i \vdot \bm{y}_j \, .
\end{equation}
Thus, the interaction term is only a small perturbation at large $a$.
We now proceed perturbatively:
to first order in $1/a$, the fixed points of \cref{eq:softspindyn} are located
at
\begin{equation}
    y_i^\mu
    =
    \left(b_i + \frac{1}{a} \eta_i\right) e_i^\mu + \order{a^{-2}} \, , 
\end{equation}
where $b_i$ can take the values $0$ or $1$, and $\eta_i = (3b_i - 1)^{-1}
\bm{e}_i \vdot \sum_j J_{ij} b_j \bm{e}_j$, with $\bm{e}_i \in
\mathbb{S}^{d-1}$ being arbitrary unit vectors.
The energy at this order in perturbation theory becomes
\begin{equation}\label{eq:energy_pert}
    a^{-2}E = -\frac{1}{4} \sum_i b_i
    - \frac{1}{2a} \sum_{ij} J_{ij} b_i b_j \bm{e}_i \vdot \bm{e}_j
    + \order{a^{-2}} \, .
\end{equation}
By computing the Hessian of the energy $E$, it can be verified that all fixed
points with some $b_i=0$ are unstable.
Hence, the local minima of $E$ have all $b_i=1$.
In that case, the second term in \cref{eq:energy_pert} is exactly a rescaled
version of the energy
\begin{equation}
    E_{\mathrm{vec}} =
	-\frac{1}{2} \sum_{ij} J_{ij} \bm{s}_i \vdot \bm{s}_j \, ,
	\quad \bm{s}_i \in \mathbb{S}^{d-1} \, ,
\end{equation}
which shows that the dynamical system in
\cref{eq:softspindyn} tends to minimize the vector spin energy at large gain.

\subsection{Dimensional reduction mechanisms}\label{sub:dimredux}

\subsubsection{Anisotropic gain}\label{sub:aniso_gain}
One option to drive the system towards Ising configurations, is to add a
quadratic anisotropy of the form $\sum_i (\bm{D}_i \vdot \bm{x}_i)^2$ that
either punishes or rewards configurations aligned with $\bm{D}$.
This corresponds to second harmonic injection-locking as implemented in complex
gain-dissipative oscillators \cite{kalinin2018global} or oscillator Ising
machines~\cite{albertsson2023HighlyReconfigurableOscillatorbased}.

If we add a quadratic anisotropy, the total energy becomes
\begin{equation}
\begin{split}
	&\phantom{=}\;\;
	E_{\mathrm{ag}}(\vb{x}, \vb{a}, \vb{D}) =
	E(\vb{x}, \vb{a}) - \frac{1}{2} \sum_i (\bm{D}_i \vdot \bm{x}_i)^2 \\ &=
	\sum_i V(\bm{x}_i, a_i)
	- \frac{1}{2} \sum_{ij} J_{ij} \bm{x}_i \vdot \bm{x}_j
	- \frac{1}{2} \sum_i (\bm{D}_i \vdot \bm{x}_i)^2 \, ,
\end{split}
\end{equation}
while the equations of motion are modified to
\begin{equation}\label{eq:eom_qa}
    \dv{t} \bm{x}_i = a_i(t) \bm{x}_i - \norm{\bm{x}_i}^2 \bm{x}_i
        + \sum_{j=1}^N J_{ij} \bm{x}_j
        + (\bm{D}_i \vdot \bm{x}_i) \bm{D}_i \, .
\end{equation}
For simplicity, we make the anisotropy the same on every site, i.e., \(\bm{D}_i
= \bm{D}\).
Since the energy \(E(\vb{x}, \vb{a})\) without the anisotropy is symmetric
under global rotations, the specific direction of \(\bm{D}\) should not matter.
Hence, we can choose $\bm{D} \sim (1, 0, \dots, 0)^\intercal$ without loss of
generality.
Then, Eq.~\eqref{eq:eom_qa} can be written
\begin{multline}\label{eq:eom_qa_1}
    \dv{t} x_i^\mu = \\
    \begin{cases}
        a_i(t) x_i^1 - \norm{\bm{x}_i}^2 x_i^1
            + \sum_{j=1}^N J_{ij} x_j^1
            + h(t) x_i^1\, , &\mu = 1 \, , \\[0.5em]
        a_i(t) x_i^\mu - \norm{\bm{x}_i}^2 x_i^\mu
            + \sum_{j=1}^N J_{ij} x_j^\mu  &\mu \neq 1 \, ,
    \end{cases}
\end{multline}
where $h = \norm{\bm{D}}^2$, and we have made explicit that the quadratic
anisotropy will generally depend on time.

Equivalently, one might absorb the quadratic anisotropy into the linear gain
term, and think about it as an anisotropic gain:
\begin{equation}\label{eq:eom_aniso_gain}
    \dv{t} \bm{x}_i = \bm{a}_i(t) \odot \bm{x}_i - \norm{\bm{x}_i}^2 \bm{x}_i
        + \sum_{j=1}^N J_{ij} \bm{x}_j \, .
\end{equation}
Here, $a^1_i = a_i + h$, $a^\mu_i = a_i$ for $\mu=2, \dots, d$, and $\odot$
denotes the element-wise product between vectors.
Written in this way, we can see that through the anisotropic gain, the first
spin component receives a larger gain relative to the others whenever \(h >
0\).
The local minima of the corresponding energy
\begin{multline}
    E_{\mathrm{ag}}(\vb{x}, \vb{a}) =
	\sum_i \left(
		-\frac{1}{2} \sum_\mu a_i^\mu (x_i^\mu)^2  + \frac{1}{4} \norm{\bm{x}_i}^4
	\right) \\
	- \frac{1}{2} \sum_{ij} J_{ij} \bm{x}_i \vdot \bm{x}_j
\end{multline}
become more and more aligned with the first component \(a^1\) as the strength
of the anisotropy \(h\) becomes larger.
Since the vector spin energy \(E_{\mathrm{vec}}\) reduces to the Ising energy
\(E_{\mathrm{Ising}}\) for collinear spins, the system thus minimizes the Ising
energy under the influence of a strongly anisotropic gain.

\subsubsection{Anisotropic metric}
Refs.~\cite{%
	calvanesestrinati2022MultidimensionalHyperspinMachine,
	calvanesestrinati2024HyperscalingCoherentHyperspin%
} proposed coupling vector spins with a time-dependent metric, referring to
this under the name of \emph{dimensional annealing}.
In our notation, this is equivalent to replacing the coupling terms as
\begin{equation}
    J_{ij} \bm{x}_i \vdot \bm{x}_j \to
	J_{ij} \sum_{\mu\nu} g^{\mu\nu}(t) x_i^\mu x_j^\nu \, ,
\end{equation}
where \(g(t)\) is a time-dependent matrix, interpreted as a metric for
\(\mathbb{R}^d\), the space of soft vector spins.
While the metric \(g(t)\) is generally determined by \(\order{d^2}\)
time-dependent coefficients, for simplicity, we restrict ourselves in the
following to diagonal metrics of the form
\begin{equation}\label{eq:gmunut}
    g^{\mu\nu}(t) =
	\begin{cases}
		1 \, , \quad &\mu=\nu = 1 \, , \\
		1 - b(t) \, , \quad &\mu=\nu \neq 1 \, , \\
		0 \, , \quad &\text{otherwise.}
	\end{cases}
\end{equation}
Here, \(b(t)\) is a monotonically increasing function mapping to the interval
\([0, 1]\) determining the dimensional annealing schedule.
The total energy associated with the anisotropic metric is
\begin{equation}\label{eq:eam}
	E_{\mathrm{am}}(\vb{x}, \vb{a}, g) = \sum_i V(\bm{x}_i, a_i) - \frac{1}{2} \sum_{ij} J_{ij} \sum_{\mu\nu} g^{\mu\nu}(t)\, x_i^\mu x_j^\nu \, .
\end{equation}
As \(b(t) \to 1\), the coupling term selectively strengthens interactions along the first
component \(x_i^1\), which biases fixed points toward first-axis alignment.
Note, however, that the on-site potential \(V(\bm{x}_i, a_i)\) retains \(O(d)\) symmetry and
does not by itself penalize transverse components.
In the limit \(b(t) \to 1\) with \(J=0\), for instance, non-axis-aligned states with
\(\norm{\bm{x}_i}=\sqrt{a}\) remain fixed points.
Introducing any finite coupling $J \neq 0$ in this limit breaks the $O(d)$ symmetry and
yields fixed points that are aligned with the first axis, thus producing Ising-like
configurations.

\subsubsection{Generalized cross product penalty}
In Ref.~\cite{cummins2025VectorIsingSpin}, the \emph{Vector Ising Spin
Annealer} (VISA) was introduced.
VISA uses three dimensional spin variables \(\bm{x}_i \in \mathbb{R}^3\) that
evolve according to the gradient of the energy (adjusted to our notation)
\begin{equation}
	E_{\mathrm{VISA}}(\vb{x}, \vb{a}, P) =
	E(\vb{x}, \vb{a}) + \frac{P}{4} \sum_{ij} \norm{\bm{x}_i \cp \bm{x}_j}^2 \, ,
\end{equation}
where \(a\) and \(P\) can depend on time, and the last term is a penalty proportional to the
norm of the cross product of all spin pairs.
Since the cross product vanishes only when two vectors are parallel, the
penalty forces spins to be collinear if \(P\) is large enough, minimizing the
Ising energy.

While the cross product cannot be straightforwardly generalized beyond \(d=3\),
it would suffice for our purposes to define a product that vanishes whenever
two vectors are collinear.
This can be done by noting that the components of the three dimensional cross
product may be written as
\begin{equation}\label{eq:cross}
	(\bm{x}_i \cp \bm{x}_j)^{\mu} =
	\sum_{\alpha\beta} \epsilon^{\mu\alpha\beta} x_i^{\alpha} x_j^\beta \, ,
\end{equation}
where \(\epsilon\) is the totally antisymmetric Levi-Civita tensor.
In \(d\) dimensions, the right-hand side of \cref{eq:cross} can be generalized to
\begin{equation}
	\sum_{\alpha\beta} \epsilon^{\mu_1 \mu_2 \dots \mu_{d-2} \alpha\beta} x_i^{\alpha} x_j^\beta \, .
\end{equation}
This is a rank \(d-2\) tensor which vanishes whenever \(\bm{x}_i = c \bm{x}_j\)
thanks to the antisymmetry of the Levi-Civita tensor.
Contracting this tensor with itself yields
\begin{equation}
\begin{split}
	&\phantom{=}\;\;
	\sum_{\mu_1 \dots \mu_{d-2}} \left(
		\sum_{\alpha\beta} \epsilon^{\mu_1 \mu_2 \dots \mu_{d-2} \alpha\beta} x_i^{\alpha} x_j^\beta
	\right)^2 \\
	&=
	\frac{(d-2)!}{2} \sum_{\alpha\beta} \left(
		x_i^\alpha x_j^\beta - x_i^\beta x_j^\alpha
	\right)^2 \\
	&=
	(d-2)! \Bigl[ \norm{\bm{x}_i}^2 \norm{\bm{x}_j}^2 - (\bm{x}_i \vdot \bm{x}_j)^2 \Bigr] \, ,
\end{split}
\end{equation}
which is the energetic penalty we are looking for.
In total, we therefore obtain
\begin{multline}\label{eq:gcpp}
	E_{\mathrm{gcpp}}(\vb{x}, \vb{a}, P) =
	E(\vb{x}, \vb{a}) \\
	+ \frac{P}{4} \sum_{ij} \Bigl[
		\norm{\bm{x}_i}^2 \norm{\bm{x}_j}^2 - (\bm{x}_i \vdot \bm{x}_j)^2
	\Bigr] \, ,
\end{multline}
which has the gradient with respect to $\bm{x}_i$
\begin{equation}
	\nabla_{\bm{x}_i} E_{\mathrm{gcpp}} = \nabla_{\bm{x}_i} E(\vb{x}, \vb{a}) + P \sum_j \Bigl(
		\norm{\bm{x}_j}^2 I  - \bm{x}_j \bm{x}_j^\intercal
	\Bigr) \bm{x}_i \, ,
\end{equation}
where \(I\) denotes the identity matrix.
Factors of \((d-2)!\) have been absorbed into the constant \(P\) in the
expressions above.

\section{Simulations and optimization performance}\label{sec:sims}
We now apply the dimensional reduction mechanisms discussed in \cref{sec:model} to perform
\emph{dimensional annealing}.
That is, using one of the dimensional reduction mechanisms, we deform the soft
vector spin energy \(E(\vb{x}, a)\) during the optimization process, biasing
the spins more and more towards collinear, \enquote{Ising-like},
configurations.
In total, we test three dimensional annealing methods:
\begin{enumerate}
	\item \emph{Anisotropic gain annealing} (AGA) uses a time-dependent anisotropic gain
		\(\bm{a}(t)\).
		The gain is initialized isotropically, i.e., \(a^\mu(0) = a_0\), and then changes
		according to a predetermined annealing schedule that affects each component
		differently.
	\item \emph{Metric annealing} (MA) uses a time-dependent metric \(g(t)\).
		The metric is initialized as the identity matrix and then changes
		according to an annealing schedule described by the function \(b(t)\)
		in \cref{eq:gmunut}.
	\item \emph{Generalized cross product penalty annealing} (GCPP) uses a
		penalty of the form in \cref{eq:gcpp}.
		The annealing schedule is given by the time-dependent coefficient of
		the penalty, \(P(t)\).
\end{enumerate}
In conjunction with the dimensional annealing, we also perform gain annealing
by varying \(a(t)\) (or \(\bm{a}(t)\) for AGA) in all methods according to a
predetermined annealing schedule.

\subsection{Annealing schedules and simulation details}
\subsubsection{Gain annealing schedules}
For the gain annealing, we use either a linear or a feedback-driven annealing
schedule.
Our linear annealing schedules are of the form
\begin{equation}
    a_i(t) = a(t) = \min\left(a_0 + \frac{t}{\tau_a}, a_{\mathrm{max}}\right) \, ,
\end{equation}
where \(a_0 = -\lmax(J)\), \(a_{\mathrm{max}} = a_0 + 2\), and \(\tau_a =
0.4t_{\mathrm{f}}\).
Here, \(t_{\mathrm{f}}\) is the total run time of the simulation and is set to
\(t_{\mathrm{f}} = 10^3\) unless otherwise specified.
The initial gain is set equal to \(-\lmax(J)\) because any smaller values would just lead to
amplitudes unnecessarily decreasing before the bifurcation (see \cref{sub:lowgain}; a
similar argument is also made in Ref.~\cite{yamamura2024GeometricLandscapeAnnealing}).
For the value of \(a_{\mathrm{max}}\), we are not aware of any simple heuristic to determine
a value of $a_{\mathrm{max}}$ a priori such that the optimization performance is superior
for any general coupling matrices.
Instead we pick the value of \(a_0 + 2\) because it leads to final soft vector
spin amplitudes that are roughly close to 1.
Lastly, the value of \(\tau_a = 0.4t_{\mathrm{f}}\) is chosen such that the
gain annealing happens over a significant fraction of the total simulation time
while still leaving plenty of time for the system to settle into a steady state
once the annealing has stopped.

The feedback-driven schedule changes the gain heterogeneously via
\begin{equation}\label{eq:fb}
    \dot{a}_i(t) =
	\epsilon (1 - \norm{\bm{x}_i}^2) \, .
\end{equation}
In order to achieve roughly similar time scales as in the linear annealing
case, we set \(\epsilon = 1/\tau_a\) (for small amplitudes, where the quadratic
term \(\norm{\bm{x}_i}^2\) can be neglected, the gain will thus grow at almost
the same rate as in the linear case).
Initial gains are set to \(a_i(0) = a_0 = -\lmax(J)\) again.
At a steady state, all amplitudes are equal, \(\norm{\bm{x}_i} = 1\).
Feedback-driven schedules thus eliminate heterogeneous amplitudes which are
usually seen as a barrier to optimization in Ising and vector spin
machines~\cite{%
	leleu2019DestabilizationLocalMinima,
	bohm2021OrderofmagnitudeDifferencesComputational,
	inui2022ControlAmplitudeHomogeneity,
	cummins2025IsingHamiltonianMinimization%
}.

Through preliminary test runs and trial-and-error, we have confirmed that the
parameters for linear and feedback-driven gain annealing parameters work
reasonably well for most instances and dimensional annealing methods though we
do not expect them to be optimal.
Later, we will investigate the effect of varying the total simulation time
\(t_{\mathrm{f}}\).

\subsubsection{Dimensional annealing schedules}
For AGA, we use the annealing schedule
\begin{equation}
    a_i^\mu(t) =
	\begin{cases}
		a_i(t) \, , \quad &\text{if } \mu=1 \, , \\
		a_i(t) - \Delta a\, b(t)  \, ,
		\quad &\text{if } \mu \neq 1 \, ,
	\end{cases}
\end{equation}
where
\begin{equation}
    b(t) =
	\begin{cases}
		0 \, , \quad &\text{if } t < t_b \, , \\
    	\min\left(\frac{t - t_b}{\tau_b}, 1\right) \, ,
		\quad &\text{if } t \geq t_b \, ,
	\end{cases}
\end{equation}
with \(t_b = 0.2t_{\mathrm{f}}\), \(\tau_b = 0.6t_{\mathrm{f}}\), and \(\Delta a =
\norm{J}_2\)$= \abs{\lmax(J)}$, where \(\norm{J}_2\) is the spectral norm.
Choosing the strength of the anisotropy, \(\Delta a\), in this way ensures that
the maximal additional energy contribution due to the anisotropy is of similar
size to the interaction energy.
In particular, we have
\begin{equation}
	\frac{1}{2} \sum_{\mu \neq 1} \abs{\sum_{ij} J_{ij} x_i^\mu x_j^\mu}
	\leq \norm{J}_2 \sum_{\mu \neq 1} \sum_j (x_j^\mu)^2 \, ,
\end{equation}
where the right-hand side is of the same order as the energy coming from the anisotropy at
large times \(t \geq t_b + \tau_b\).
The start time, \(t_b\), is chosen such that the system has enough time to
settle into a low-energy vector spin state before the dimensional annealing
kicks in (cf., \cref{fig:en_evol}).
Similarly to \(\tau_a\), \(\tau_b\) is chosen such that the dimensional
annealing takes place over a significant fraction of the total run time, while
leaving enough time for the system to settle into a steady state at the end.
Our MA simulations use the same \(b(t)\) as in AGA,
with the same values for \(t_b\) and \(\tau_b\).
Lastly, for GCPP annealing, we use
\begin{equation}
    P(t) = P_{\mathrm{max}}\, b(t)
\end{equation}
where \(b(t)\) again stays the same, while \(P_{\mathrm{max}} = \norm{J}_2 /
N\).
The heuristic for setting \(P_{\mathrm{max}}\) to this values follows a similar
logic to that of \(\Delta a\) in AGA.
Assuming \(\norm{\bm{x}_i}^2 \norm{\bm{x}_j}^2 - (\bm{x}_i \vdot \bm{x}_j)^2
\approx 1\), the maximal penalty term contribution to the energy is of roughly
similar size to the interaction term this way.


\subsubsection{Numerical integration and example trajectories}
In all cases we integrate the equation of motion
\begin{equation}
	\dv{\vb{x}}{t} =
	-\bm{\nabla} E_{\mathrm{da}}
\end{equation}
(in conjunction with \cref{eq:fb} in the case of feedback-driven schedules) in
a time interval \([0, t_{\mathrm{f}}]\) using SciPy's \texttt{solve\_ivp}
function~\cite{virtanen2020SciPy10Fundamental} which uses an adaptive RK45
method~\cite{dormand1980FamilyEmbeddedRungeKutta} internally.
The default tolerances are used (\(10^{-3}\) and \(10^{-6}\) for relative and absolute
tolerances, respectively).
In the above, the energy \(E_{\mathrm{da}} = E_{\mathrm{ag}}, E_{\mathrm{am}},
E_{\mathrm{gcpp}}\) is the energy associated with the dimensional annealing
method.
Initial conditions are always sampled from the uniform distribution on \([-0.1,
0.1]^{N \times d}\).
Additionally, we normalize all coupling matrices \(J\) by a scale factor such
that the Ising ground state energies are roughly between \(-1\) and \(-10\).
This is done mainly to ensure that the gradient of the interaction energy does
not become very large which would require very small step sizes in the
numerical integration.
The scale factors for different coupling matrices are shown in
\cref{tab:scale_factors}.
We perform simulations for soft vector spin dimensions ranging from \(d=1\) to
\(d=5\).
Note that for \(d=1\), dimensional reduction is not possible, and we only use
gain annealing.
Nevertheless, we include the \(d=1\) simulations as a baseline to compare the
dimensional annealing methods for \(d>1\) with.

\cref{fig:en_evol} shows the evolution of \(E_{\mathrm{Ising}}\) and
\(E_{\mathrm{vec}}\) for different dimensional annealing methods in \(d=3\) and
\(d=1\).
Note that for this example, \(t_{\mathrm{f}} = 2 \times 10^{3}\), and
feedback-driven gain annealing has been used.
After an initial transient, the soft vector spins reach a minimum of the vector
spin energy.
When dimensional annealing starts, the energy is driven upwards again to settle
into the minimum of the Ising energy.
Projected to Ising spins (see \cref{sub:proj}), this looks like a series of spin flips that leads to
jumps in the Ising energy.
In comparison, the \(d=1\) soft Ising spins quickly settle into a local minimum
of the Ising energy from which they are unable to escape.

\begin{figure*}
	\centering
	\includegraphics[width=6in]{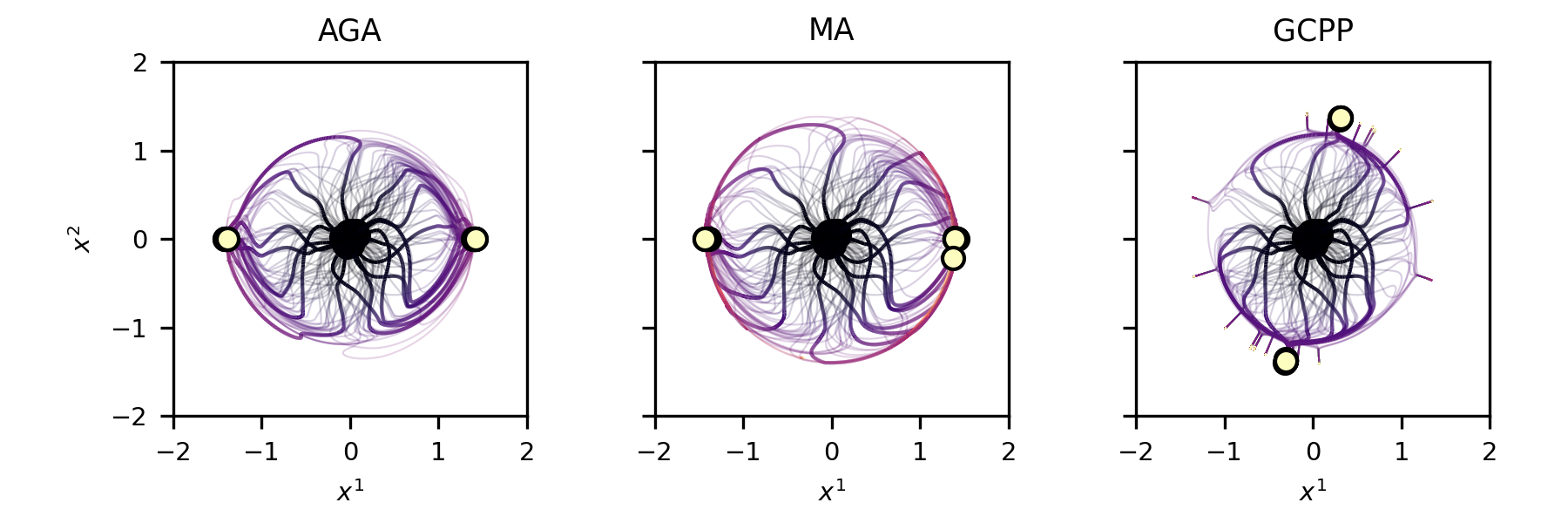}
	\caption{%
		Trajectories in a \(d=2, N=16\) soft vector spin system for different
		dimensional annealing methods and linear annealing schedules.
		Note that \(t_{\mathrm{f}} = 2 \times 10^3\) in this case.
		The couplings matrix in this case comes from the WPE.
		Each line shows the trajectory of a single spin with lighter colours
		signifying increasing times.
		Faded out lines in the background show trajectories obtained from other
		initial conditions and the same coupling matrix.
		Initial and final states are marked with black and pale yellow dots
		respectively.
		Note also that the set of initial conditions is the same for all three
		methods.%
	}
	\label{fig:traj2d}
\end{figure*}
\cref{fig:traj2d} shows example trajectories for soft vector spins in \(d=2\).
Evidently, the different dimensional annealing methods display quite distinct
dynamical behaviour.
The final states in both AGA and MA are aligned in the \(x^1\)-direction, while
in GCPP, the final states are aligned along a random axis.
That is because \(E_{\mathrm{gcpp}}\) retains the \(O(d)\) symmetry of the soft
vector spin energy \(E\), and the symmetry is spontaneously broken during
time-evolution.
The additional dimensional reduction terms in \(E_{\mathrm{ag}}\) and \(E_{\mathrm{am}}\),
on the other hand, break the \(O(d)\) symmetry explicitly by making either gain or metric
explicitly anisotropic.

\subsection{Problem classes}
We sample coupling matrices \(J\) from a number of different problem classes:
the 2D and 3D \emph{tile planted ensemble} (TPE)~\cite{%
	hamze2018EternitySpinglassPlanting,
	perera2020ComputationalHardnessSpinglass%
}, the \emph{Wishart planted ensemble}
(WPE)~\cite{hamze2020WishartPlantedEnsemble},
and \emph{random sparse matrices}.
The reason we choose these problems is that they represent a wide range of
possible coupling matrices (local vs.\ all-to-all, structured vs.\
unstructured, easy vs.\ hard, etc.) which allows us to generalize our findings
with greater confidence.
Furthermore, the planted ensembles are useful because their exact ground states
are always known.
In the case of random sparse matrices, ground states have been previously
computed through extensive Monte Carlo search.

Matrices from the 2D-TPE describe a square lattice graph based on four families
of frustrated unit-cells.
Depending on the frequency with which the different unit cell families occur,
the 2D-TPE can have a varying computational hardness.
Similarly, the 3D-TPE describes a cubic lattice graph consisting of unit cells
coming from six different families with varying hardness depending on the
distribution of unit cell families within the total graph.
Crucially, in both the 2D and 3D case, the lattice is constructed in such a way
that the fully ferromagnetic state \(\{s_i = 1\}_{i=1,\dots N}\) is always the
ground state.

The WPE is a kind of anti-Hopfield model~\cite{nokura1998SpinGlassStates} with
a number \(M=\alpha N\) of patterns that are correlated in such a way to
guarantee that the fully ferromagnetic state is always the ground state.
Varying the number of patterns \(M\) drives the WPE through a easy-hard-easy
transition with the hardest problems occurring at \(M^* \approx 1.63 +
0.073N\)~\cite{hamze2020WishartPlantedEnsemble}.

We use the library \texttt{Chook}~\cite{perera2021ChookComprehensiveSuite} to
sample instances each of 2D-TPE, 3D-TPE, and WPE coupling matrices for varying
difficulties (see \cref{tab:scale_factors}).
For each difficulty and problem class, we generate 100 different instances.
\texttt{Chook} automatically transforms all coupling matrices via
\(\mathbb{Z}_2\) gauge transformations to yield random ground states instead of
just the ferromagnetic one.
The WPE matrices use a slightly modified formulation, where all matrix entries
are integers.
This helps to avoid any rounding errors when comparing exact ground states.
Note also that the size \(N=16\) is quite small for WPE instances.
This is because even the \enquote{easy} instances appear to already be quite
hard for soft vector spin solvers.
Random sparse matrices are as in
Ref.~\cite{calvanesestrinati2024HyperscalingCoherentHyperspin}: each matrix
element is nonzero with probability \(0.2\), with the value of the nonzero
elements being \(\pm J_0\) with equal probability.
In our case, \(J_0 = 0.03\).
\begin{table*}
    \centering
    \caption{%
		Scale factors multiplying the coupling matrix and ensemble parameters
		governing difficulty for different problem classes.
		The ensemble parameters in \texttt{Chook} for TPE and WPE have been
		chosen according to the hardness phase transitions predicted in
		Refs.~\cite{%
			hamze2018EternitySpinglassPlanting,
			perera2020ComputationalHardnessSpinglass,
			hamze2020WishartPlantedEnsemble%
		}.
		In the case of TPE instances, they determine the probabilities of the
		various unit cell families being generated in the graph.
		For WPE, there is just one parameter that determines the number of
		correlated patterns \(M = \alpha N\).%
	}
    \begin{ruledtabular}
    \begin{tabular}{llll}
		problem class & size & scale factor & ensemble parameters \\ \midrule
		2D TPE & 8 $\times$ 8 & 0.02 & \(p_1 = 0.1\), \(p_2 = 0.0\), \(p_3 = 0.9\) (easy) \\
		 & & & \(p_1 = 0.5\), \(p_2 = 0.4\), \(p_3 = 0.1\) (medium) \\
		 & & & \(p_1 = 0.05\), \(p_2 = 0.9\), \(p_3 = 0.05\) (hard) \\
		  3D TPE & 4 $\times$ 4 $\times$ 4 & 0.02 & \(p_{F22} = p_{F42} = 0.4 \) (easy) \\
		 & & & \(p_{F22} = p_{F42} = 0.2 \) (medium) \\
		 & & & \(p_{F22} = p_{F42} = 0.0 \) (hard) \\
		  WPE easy & 16 & $2 \times 10^{-5}$ & \(\alpha = 0.88\) \\
		  WPE hard & 16 & $1 \times 10^{-4}$ & \(\alpha = 0.19 \approx M^* / N\) \\
		  random sparse & 30 & 1.0  & n/a \\
						& 40 & 1.0 & n/a \\
    \end{tabular}
    \end{ruledtabular}
    \label{tab:scale_factors}
\end{table*}

\subsection{Projection of soft vector to Ising spins}
\label{sub:proj}
After integrating the equations of motion we are left with a final soft vector spin state
\begin{equation}
    \vb{x}_{\mathrm{f}}
	=
	\vb{x}(t_{\mathrm{f}})
	= (\bm{x}_1(t_{\mathrm{f}}), \dots, \bm{x}_N(t_{\mathrm{f}})) \, ,
\end{equation}
which still has to be converted to an Ising state \(\vb{s}_{\mathrm{f}} \in
\{-1, 1\}^N\) to be inserted into the Ising energy, \(E_{\mathrm{Ising}}\).
There are multiple ways to do this.
For AGA and MA, it would suffice to just project to the easy axis defined by
the gain or metric anisotropy (the first axis in our case), i.e.,
\begin{equation}
    \vb{s}_{\mathrm{f}}
	=
	(\sign(\bm{x}_1^1), \dots, \sign(\bm{x}_N^1)) \, .
\end{equation}
In the case of GCPP, however, we do not know \emph{a priori} which axis the
system will align with at the fixed point.
Hence, we generally use an averaging procedure to decide the projection axis.

The procedure works as follows:
For each site \(i\), we compute \(q_i = \bm{x}_i \vdot \bm{x}_1\).
Based on that, we compute the state
\begin{equation}
    \vb{\tilde{x}}_{\mathrm{f}}
	= (\bm{\tilde{x}}_1, \dots, \bm{\tilde{x}}_N) \, ,
\end{equation}
where
\begin{equation}
    \bm{\tilde{x}}_i
	=
	\begin{cases}
		\phantom{-}\bm{x}_i \, ,\quad& q_i \geq 0 \\
		-\bm{x}_i \, ,\quad& \text{otherwise.}
	\end{cases}
\end{equation}
The state \(\vb{\tilde{x}}_{\mathrm{f}}\) has each spin aligned along the same
axis as the corresponding spin in \(\vb{x}_{\mathrm{f}}\), but has the
additional property that all spins live in a common hemisphere.
Next, we average all these spins and obtain
\begin{equation}
    \bm{x}_{\mathrm{avg}}
	=
	\frac{1}{N} \sum_{i=1}^N \bm{\tilde{x}}_i
\end{equation}
If the norm of \(\bm{x}_{\mathrm{avg}}\) is sufficiently large (greater than \(10^{-5}\) in our case), we take
\begin{equation}
	\bm{\hat{\pi}} = \frac{\bm{x}_{\mathrm{avg}}}{\norm{\bm{x}_{\mathrm{avg}}}}
\end{equation}
as the projection vector, such that
\begin{equation}
    \vb{s}_{\mathrm{f}}
	=
	(\sign(\bm{\hat{\pi}} \vdot \bm{x}_1), \dots, \sign(\bm{\hat{\pi}} \vdot \bm{x}_N)) \, ,
\end{equation}
where we use the convention that \(\sign(0) = 1\) to decide ambiguous cases.
If the norm of \(\bm{x}_{\mathrm{avg}}\) is too small, which usually happens if
the spins are either not aligned enough or amplitudes are too small, we simply
take
\begin{equation}
	\vb{s}_{\mathrm{f}}
	=
	(\sign(\bm{x}_1^1), \dots, \sign(\bm{x}_N^1)) \, .
\end{equation}
However, it is quite rare to end up in this case as long as the annealing
parameters are properly chosen.
Finally we note that more efficient and robust methods to find a good
projection axis probably exist, but the simple procedure outlined above
suffices for our purposes.
In particular, for our annealing parameters, we usually find final configurations that are
already strongly aligned along a single axis.
In that regime, the particulars of the projection method should not matter much as long as
the axis of alignment is correctly identified.

\subsection{Comparison of annealing schemes for different problem classes and dimensions}
\begin{figure*}
	\centering
	\includegraphics[width=\linewidth]{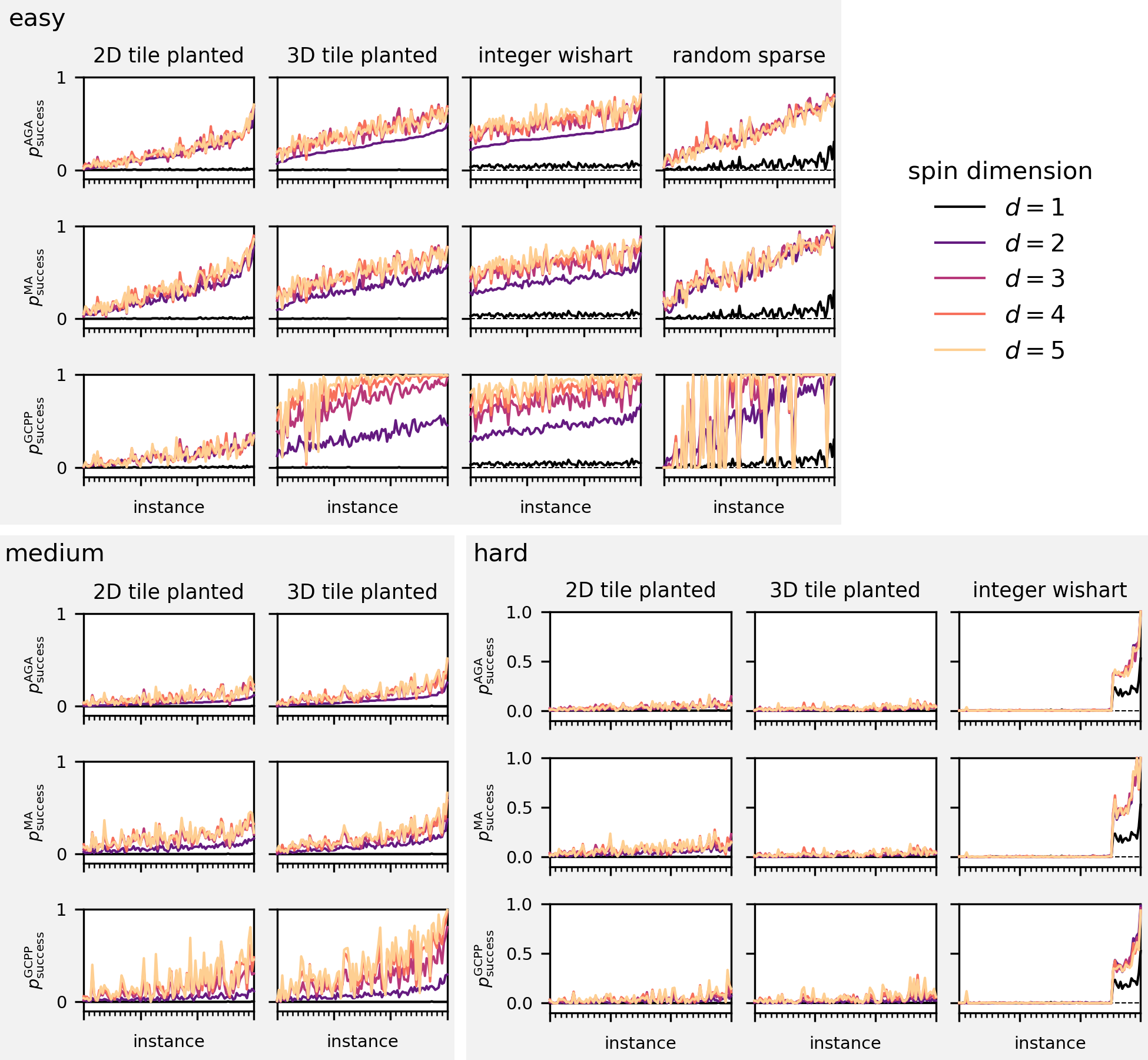}
	\caption{%
		Success probabilities for finding ground states of \enquote{easy},
		\enquote{medium}, and \enquote{hard} problems using soft vector spins
		in dimensions \(d=1,\dots,5\) and linear annealing schedules.
		The horizontal axes show the 100 different instances of each problem
		class, while the vertical axes show the success probability for
		different dimensional annealing methods.
        To aid visualization, instances are sorted on the horizontal axis in order of the success rate of $d=2$ AGA.%
	}
	\label{fig:lin_instance}
\end{figure*}
\begin{figure*}
	\centering
	\includegraphics[width=\linewidth]{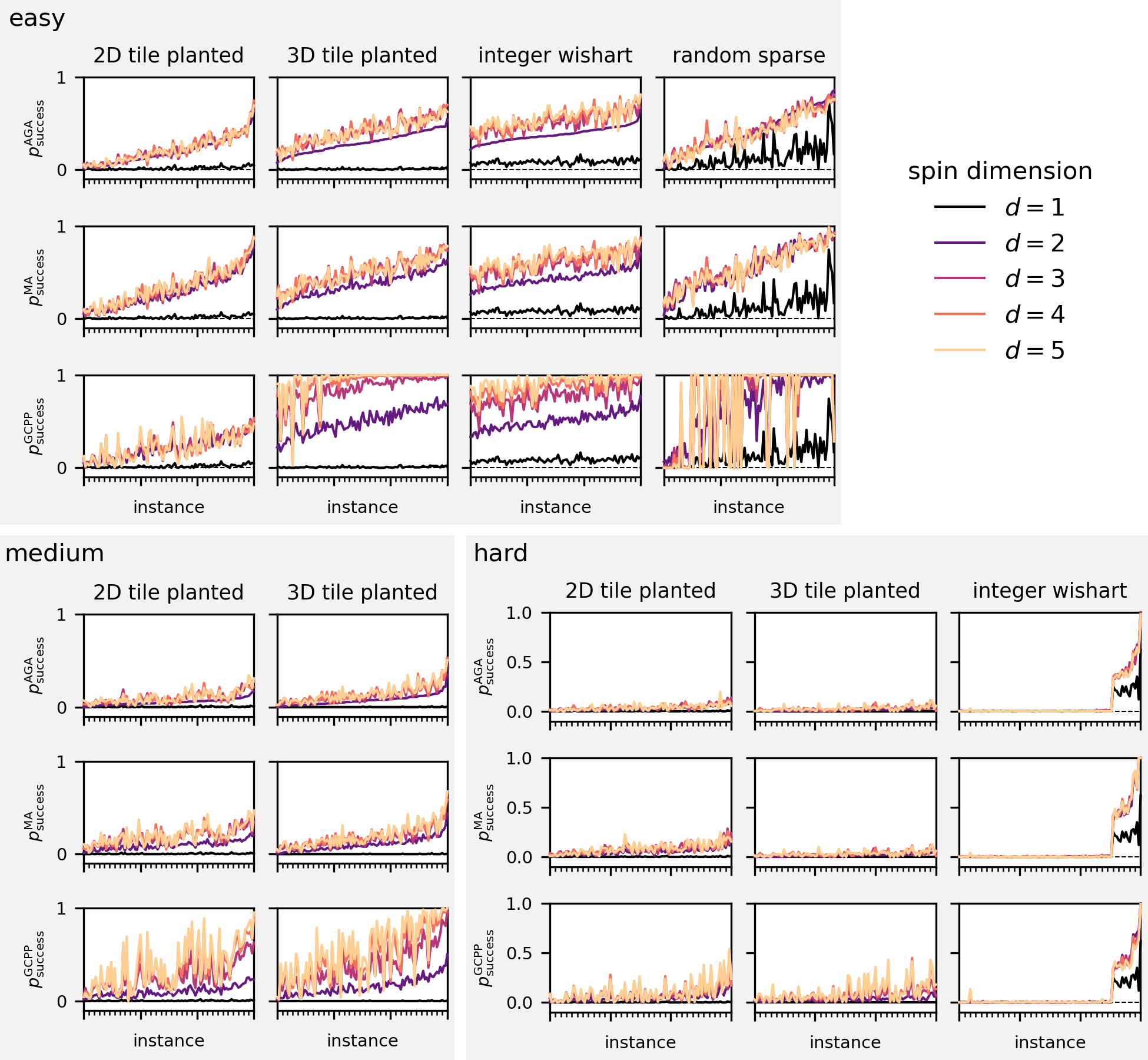}
	\caption{%
		Success probabilities for finding ground states of \enquote{easy},
		\enquote{medium}, and \enquote{hard} problems using soft vector spins
		in dimensions \(d=1,\dots,5\) and a feedback mechanism to control
		gains.
		The horizontal axes show the 100 different instances of each problem
		class, while the vertical axes show the success probability for
		different dimensional annealing methods.
        To aid visualization, instances are sorted on the horizontal axis in order of the success rate of $d=2$ AGA.%
	}
	\label{fig:fb_instance}
\end{figure*}
In \cref{fig:lin_instance,fig:fb_instance}, we compare the success
probabilities of finding the ground state for different problem classes,
difficulties, dimensional annealing methods, and soft vector spin
dimensionalities.
The success probabilty here is calculated via
\begin{equation}
    p_{\mathrm{success}} =
	\frac{\# (\text{final states with \(E_{\mathrm{Ising}} \approx E_0\)})}{N_{\mathrm{init}}} \, ,
\end{equation}
where \(E_0\) is the known ground state energy of the given coupling matrix,
and \(N_{\mathrm{init}}\) is the number of different initial conditions for the
time evolution.
In all cases, we chose \(N_{\mathrm{init}}=200\).
We define \(E_{\mathrm{Ising}} \approx E_0\) to be true if they do not differ
within an absolute tolerance of \(10^{-5}\) and a relative tolerance of \(5
\times 10^{-3}\).

Generally, we find that soft vector spins with \(d > 1\) have a higher
probability of finding the ground state than one-dimensional soft spins
(\(d=1\), i.e., ordinary Ising machines).
In almost all cases, \(p_{\mathrm{success}} \approx 0\) for \(d=1\) for the
problem classes and parameters we tested (the only exception being some of the
hard WPE instances and random sparse matrices).
The largest jump in success probability usually occurs going from \(d=1\) to
\(d=2\), with a slightly smaller jump going from \(d=2\) to \(d>2\).
Dimensionality does not seem to strongly affect the success probability beyond
\(d > 2\).

At this point, we should also note that our categorization into \enquote{easy},
\enquote{medium}, and \enquote{hard} problems is quite a rough estimate of
actual complexity.
In particular, \enquote{easy} problems are still not trivial in many cases for
the optimization methods under consideration; \enquote{medium} problems often
have sub-percentage success probabilities, and \enquote{hard} problems seem
often to be impossible for our soft vector spin solvers.
Recently, it has been shown~\cite{ghimenti2025GeometryDynamicsAnnealed} that
recovering the planted solution of the WPE is in fact impossible for small
\(\alpha\) and \(d=1\).
Another notable observation is that roughly a sixth of \enquote{WPE} instances
have actually very high success probabilities for all tested methods.
We conjecture that this could be a finite-size effect, and that the location
and sharpness of the hardness phase transition in the WPE is affected by the
quite small size (\(N=16\)) of our WPE instances.

While qualitatively similar in most respects, the results for the various
dimensional annealing methods do exhibit some minor differences.
AGA and MA have roughly similar success probabilities, while GCPP's tend to be
slightly higher on average.
However, this is not consistent across problem classes and the relatively small
differences might vanish for slightly different parameters.

Comparing the results in \cref{fig:lin_instance} with those in
\cref{fig:fb_instance}, we find that the feedback-driven approach generally
produces higher success probabilities across the board.
This is to be expected due to its ability to avoid amplitude heterogeneity.
The qualitative differences between dimensional annealing methods in different
dimensions remain roughly the same as in the linear case.
Our results thus seem to be robust against some variations in the gain
annealing schedule.

\begin{figure*}
	\centering
	\includegraphics[width=\linewidth]{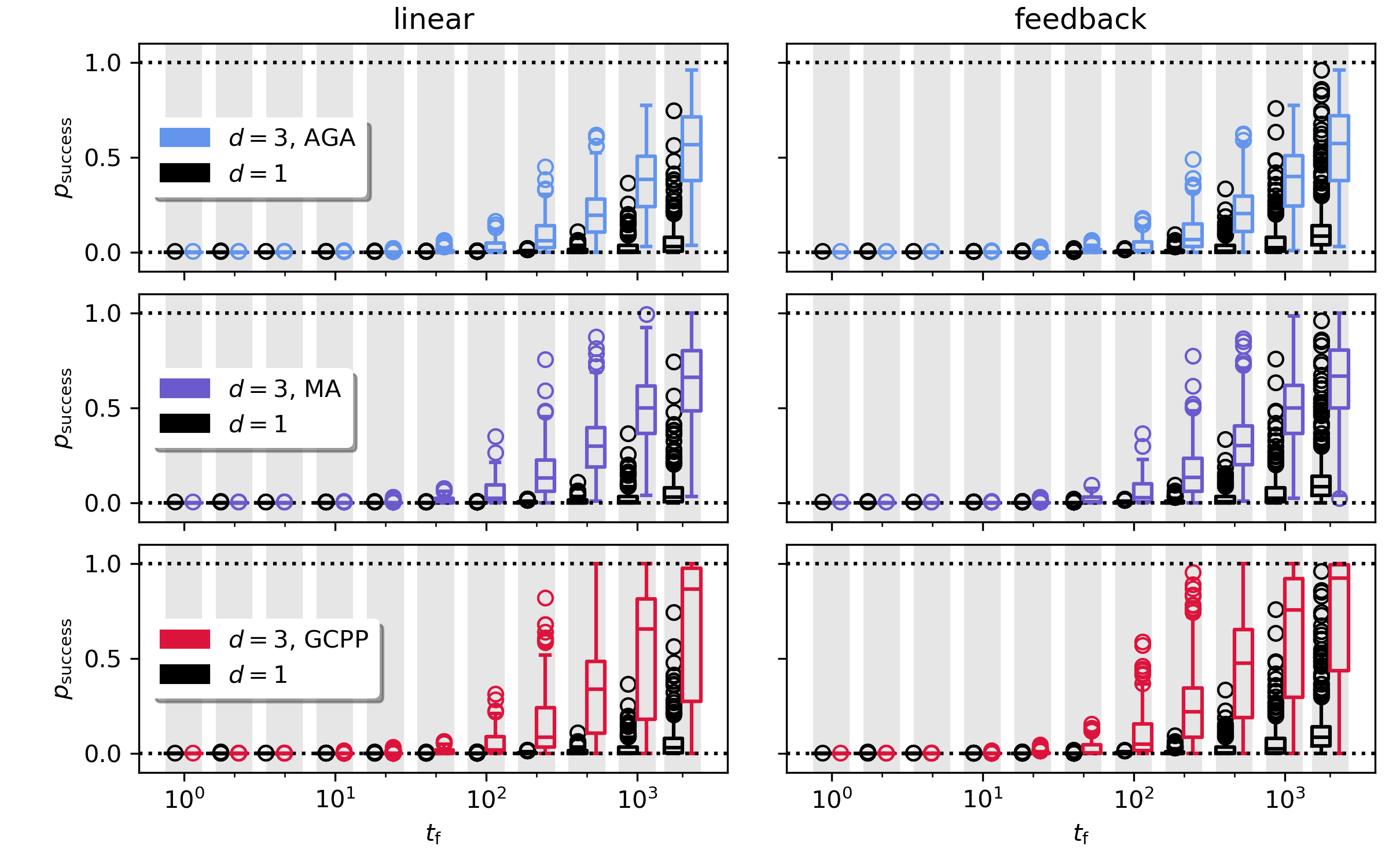}
	\caption{%
		Scaling of success probabilities with \(t_{\mathrm{f}}\) on \enquote{easy} problems.
		We calculate the success probability over 200 different initializations
		of each method for 11 values of \(t_{\mathrm{f}}\) ranging from \(1\)
		to \(2 \times 10^3\), comparing AGA, MA, and GCPP in \(d=3\) dimensions
		to just performing gain annealing in \(d=1\) dimension.
		All of this is done for both linear and feedback-driven gain annealing
		schedules.%
	}
	\label{fig:tf}
\end{figure*}
We further confirm the robustness with regards to variations in parameters and
annealing schedules by checking how the success probabilities change when
\(t_{\mathrm{f}}\) is varied.
To reduce the amount of simulations needed, we only compare \(d=3\) and
\(d=1\), and only on the \enquote{easy} problems.
The results are displayed in \cref{fig:tf}.
As one might expect, the success probabilities generally increase when the
total run time \(t_{\mathrm{f}}\) is increased.
As with our previous results, there is a clear improvement in success
probability for \(d=3\) compared to \(d=1\) for all values of
\(t_{\mathrm{f}}\) we tested.
Similarly, feedback-driven gain annealing again shows an advantage over linear
schedules.

\section{Discussion}\label{sec:disc}
We have shown how soft vector spin systems with dimensional annealing can be
used as combinatorial optimization solvers for finding the ground state of the
Ising model.
In particular, we have analyzed the dependence of dimensional annealing on
vector spin dimension for three different dimensional reduction methods based
on an anisotropic gain, an anisotropic metric, and a generalized cross product
penalty respectively.
For all these methods, and across different classes of coupling matrices with
varying hardness levels, we have shown that multi-dimensional soft vector spins
fare better at finding the ground state than their one dimensional counterpart.
More specifically, there seems to be a distinct jump in the probability of finding the
ground state when going from $d=1$ to $d=2$ and from $d=2$ to $d=3$.
Increasing the dimension beyond that yields diminishing returns.
We note also, however, that for the hardest problem instances, success probabilities remain
near zero even with dimensional annealing, indicating that the approach primarily benefits
easy-to-medium difficulty problems in the regimes we tested.
Of course, this advantage comes with a cost: on an ordinary computer it takes
more computational resources to simulate higher dimensional soft vector spins.
However, this cost could be absent (or at least lower) when analog hardware is
used.
In fact, many analog hardware platforms already have the capability of
implementing \(d=2\) soft vector spins.
That is because many Ising machines consist of analog units with 2D
amplitude and phase degrees of freedom; this includes coherent laser
light~\cite{%
	wang2013CoherentIsingMachine,
	kim2024CombinatorialClusteringCoherent%
},
polariton condensates~\cite{%
	berloff2017RealizingClassicalXY,
	lagoudakis2017PolaritonGraphSimulator%
}, or electronic
oscillators~\cite{albertsson2023HighlyReconfigurableOscillatorbased}.
We have thus shown that significant performance improvements could be achieved
in such platforms by implementing dimensional annealing.
Let us also note that recent work~\cite{kimAcceleratingCoherentIsing2026} successfully
implemented a highly optimized version of $d=2$ AGA in nondegenerate optical
parametric oscillators---further corroborating our results.

On top of that, recent work~\cite{%
	calvanesestrinati2022MultidimensionalHyperspinMachine,
	calvanesestrinati2024HyperscalingCoherentHyperspin,
	chiavazzo2025IsingMachineDimensional%
} has made progress in finding analog hardware capable of implementing higher
dimensional spins with \(d \geq 2\).
In particular Refs.~\cite{
	calvanesestrinati2022MultidimensionalHyperspinMachine,
	calvanesestrinati2024HyperscalingCoherentHyperspin%
} introduced dimensional annealing through anisotropic metrics in the context
of the hyperspin machine.
We have expanded on their work by generalizing the notion of dimensional
annealing to include different dimensional reduction methods as well as
applying it to a more abstract and general model of soft vector spins.

An obvious question is the implementability of the different dimensional
reduction mechanisms and dimensional annealing methods in hardware. The answer
to this question is rather straightforward in the case of AGA: as mentioned in
\cref{sub:aniso_gain}, anisotropic gain corresponds directly to second-harmonic
injection locking.
As outlined in Ref.~\cite{calvanesestrinati2024HyperscalingCoherentHyperspin},
MA can be implemented using parametric oscillators, either in an all-optical
setup (\(d=2\)), or in an opto-electronic hybrid approach (\(d>2\)).
GCPP might be the most challenging of the methods to implement, due to the
penalty term representing a 4-body interaction of the soft vector spins.
This leads to an additional nonlinear coupling of the spins which suitable
hardware platforms would have to be capable of implementing.
Such nonlinear couplings may be realized in integrated photonic circuits or
free space optics via nonlinear optical effects, or outsourced to electronic
hardware in a hybrid setup, as argued in
Ref.~\cite{cummins2025VectorIsingSpin}.

When Refs.~\cite{%
	calvanesestrinati2022MultidimensionalHyperspinMachine,
	calvanesestrinati2024HyperscalingCoherentHyperspin%
} introduced dimensional annealing, showing how it can help optimization, they
did so with fixed gains and sparse random coupling matrices.
We have now verified that their results also hold for dynamically varying
gains, either with homogeneous linear annealing schedules, or heterogeneous
feedback-driven annealing, and with a wide range of different coupling
matrices.
Furthermore, we have shown how the advantage of dimensional annealing persists
even when the total run time increases.
In total, this yields even more robust evidence that the use of vectorial
degrees of freedom offers an advantage for analogue optimization solvers.

That being said, more extensive benchmarking might be useful for identifying
optimal parameters and annealing schedules or how to adapt these to specific
coupling matrices.
More advanced feedback-based annealing schemes like chaotic amplitude control~\cite{%
	leleu2019DestabilizationLocalMinima,
	leleu2021ScalingAdvantageChaotic%
} would also be worthwhile to examine in conjunction with dimensional
annealing.
Additionally, benchmarking against established solvers such as simulated annealing or
parallel tempering would help to contextualize the absolute performance of soft vector spin
methods beyond the relative comparison between $d=1$ and $d>1$ presented here.
Apart from that, theoretical calculations could provide hints on how to
optimize the performance of soft vector spin systems.
Previously, studies of the high-dimensional energy landscape of Ising
machines~\cite{%
	yamamura2024GeometricLandscapeAnnealing,
	ghimenti2025GeometryDynamicsAnnealed%
} have proven useful to find optimal annealing schedules in the \(d=1\) case.
Soft vector spins might also be intriguing to study removed from the context of
optimization.
For purely ferromagnetic couplings, the soft vector spin model represents a
discretized version of the \(O(d)\) field theory, relevant for quantum and
classical critical phenomena~\cite{zinn-justin2021QuantumFieldTheory}.
Our model thus represents a generalization of this model with disordered
couplings.
We hope that our work can inspire further research into the dynamics of soft
vector spins as combinatorial optimization solvers specifically and as a
complex system generally.

\begin{acknowledgments}
	The authors wish to thank Marcello Calvanese Strinati for useful discussions as well as
	for providing the dataset of random sparse coupling matrices and their ground states.
	The authors acknowledge the support from HORIZON EIC-2022-PATHFINDERCHALLENGES-01
	HEISINGBERG Project 101114978.
    M.S.\ acknowledges support from EPSRC (grant UKRI2897).
	R.Z.W.\ and N.G.B.\ acknowledge the support from Julian Schwinger Foundation Grant
	No.~JSF-19-02-0005. 
	N.G.B.\ also acknowledges support from Weizmann-UK Make Connection Grant 142568 and the
	EPSRC UK Multidisciplinary Centre for Neuromorphic Computing (grant UKRI982).
\end{acknowledgments}

\bibliography{refs}

\end{document}